\documentclass[aps,prl,reprint,superscriptaddress,nobibnotes]{revtex4-2}

\usepackage{amsmath,amssymb,amsthm,mathtools}
\usepackage{bm}
\usepackage{bbm}
\usepackage{physics}
\usepackage{xcolor} 
\usepackage{times}
\usepackage[hypertexnames=false,colorlinks=true,citecolor=teal,linkcolor=teal,urlcolor=teal]{hyperref}

\newcommand{\STAB}{\mathrm{STAB}}
\newcommand{\PPT}{\mathrm{PPT}}
\newcommand{\cone}{\operatorname{cone}}
\newcommand{\conv}{\operatorname{conv}}
\newcommand{\Ad}{\mathrm{Ad}}
\newcommand{\Cl}{\mathrm{Cl}}

\newtheorem{theorem}{Theorem}

\newtheorem{proposition}[theorem]{Proposition}
\newtheorem{corollary}[theorem]{Corollary}
\newtheorem{conjecture}[theorem]{Conjecture}

\theoremstyle{definition}
\newtheorem{definition}{Definition}

\begin{document}

\title{Bound magic in multi-qubit  systems}

\author{Jonas Kitzinger}
\email{jonas.kitzinger@fu-berlin.de}
\affiliation{Dahlem Center for Complex Quantum Systems, Freie Universit{\"a}t Berlin, 14195 Berlin, Germany}

\author{Jens Eisert}
\affiliation{Dahlem Center for Complex Quantum Systems, Freie Universit{\"a}t Berlin, 14195 Berlin, Germany}
\affiliation{Helmholtz-Zentrum Berlin f{\"u}r Materialien und Energie, 14109 Berlin, Germany}
\affiliation{Fraunhofer Heinrich Hertz Institute, 10587 Berlin, Germany}

\author{Ravi Kunjwal}
\affiliation{Aix-Marseille University, CNRS, LIS, Marseille, France}

\date{\today}

\begin{abstract}
Magic captures nonstabilizerness, a resource required for quantum computational speedups. 
We prove the existence of multi-qubit nonstabilizer states from which no pure magic can be distilled, even with arbitrarily many copies and rounds of stabilizer operations. 
Such bound magic states have previously been only known to exist in odd Hilbert-space dimensions, where positivity of discrete Wigner functions can be used analogously to the positive partial transpose (PPT) 
criterion for bound entanglement. 
We introduce a PPT analogue for magic in multi-qubit systems that is closed under tensor products and stabilizer operations and reduces to the stabilizer polytope for one and two qubits. 
This enables the proof of multi-qubit bound magic and the explicit construction of bound magic states for $n \geq 3$ qubits.
\end{abstract}

\maketitle

Quantum resource theories provide a precise framework for identifying which quantum states are useful under a restricted set of operations~\cite{chitambar2019QuantumResourceTheories}. 
They divide quantum state space into free states and resourceful states, specified either axiomatically or through a set of free operations.
In the paradigmatic resource theory of entanglement, the free states are separable, resourceful states are entangled, and free operations are \emph{local operations and classical communication} (LOCC).
In the resource theory of magic~\cite{veitch2014ResourceTheoryStabilizer}, which will be the focus of this Letter, the free states are mixtures of stabilizer states, resourceful states are magic~\cite{bravyi2005UniversalQuantumComputation} (or nonstabilizer) states, and the set of stabilizer operations assumes the role of LOCC, capturing a notion of ``quantumness'' of states in quantum physics, originally as resource states in the magic-state injection model of quantum computation, and more recently also within quantum many-body theory \cite{oliviero2022MagicstateResourceTheory,niroula2024PhaseTransitionMagic,sierant2026TheoryMagicPhase,white2026ExactOperatorComplexity}.

The study of magic as a resource is, in part, a foundational question: it asks which quantum states lie beyond the stabilizer description, and what this distinction means operationally. However, it also has a concrete role in fault-tolerant quantum computing. 
Stabilizer states and operations alone admit efficient classical simulation~\cite{aaronson2004ImprovedSimulationStabilizer}.
Magic states supply the non-Clifford ingredient needed for computational universality. 
Magic-state distillation turns suitable noisy states into high-fidelity resources~\cite{bravyi2005UniversalQuantumComputation,litinski2019MagicStateDistillation}, while magic-state cultivation grows a reliable encoded magic state from an initial injection~\cite{gidney2024MagicStateCultivation}.
These states can then be injected into fault-tolerant Clifford circuits to implement non-Clifford gates.
This makes it natural to ask whether every nonstabilizer state can ultimately be distilled into pure magic, and it is a relevant question for the current efforts towards fully fault-tolerant quantum computers~\cite{eisert2026MindGapsFraught}.

A celebrated early result in entanglement theory has shown that the question of resource distillation and the division into free and resourceful states is more nuanced than one might initially expect: There exist nonseparable states that require entanglement for their preparation, but from which no pure entangled states can be distilled using LOCC~\cite{horodecki1998MixedstateEntanglementDistillation}. 
Such undistillable nonseparable states are known as bound entangled states. 
The \emph{positive partial transpose} (PPT) criterion~\cite{peres1996SeparabilityCriterionDensity,horodecki1996SeparabilityMixedStates} has turned out to be crucial in this distinction because every nonseparable PPT state is bound entangled. 

\begin{figure}
    \centering
    \includegraphics[width=\linewidth]{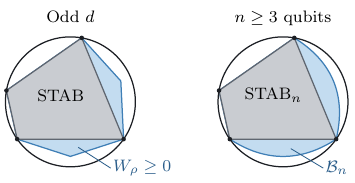}
    \caption{Qualitative sketch of the stabilizer polytope $\STAB$ surrounded by the set of bound magic states inside quantum state space. The vertices represent pure stabilizer states. The left-hand side shows the situation in odd local Hilbert-space dimensions, where the blue region represents the region $W_\rho^+$ of states $\rho$ with positive Wigner function $W_\rho$. It is strictly larger than $\STAB$, and states in $W_\rho^+ \setminus \STAB$ are bound magic. The right-hand side shows the corresponding situation in multi-qubit Hilbert space, where the set $\mathcal{B}_n$ assumes the role of Wigner-positive states.}
    \label{figure}
\end{figure}

Whether an analogous phenomenon to bound entanglement exists for magic has been one of the first questions asked in the resource theory of magic~\cite{bravyi2005UniversalQuantumComputation,veitch2014ResourceTheoryStabilizer}. In odd dimensions, Gross' discrete Wigner function~\cite{gross2006HudsonTheoremFinitedimensional} can play the role analogous to PPT and provides a path to prove that bound magic states exist~\cite{veitch2012NegativeQuasiprobabilityResource}. Similarly to PPT entangled states, all known bound magic states are nonstabilizer but Wigner-positive.

For even dimensions, however, including the multi-qubit systems most commonly used in quantum computing theory, no Wigner function is known that has the properties required for a proof of bound magic analogous to the odd-dimensional case~\cite{zhu2016PermutationSymmetryDetermines}.
Previous no-go results resembling bound magic for qubits were limited statements, either restricting to finite-length protocols on single-qubit states as in Ref.~\cite{campbell2010BoundStatesMagic}, looking at different tasks such as exact purification of full-rank resource states~\cite{fang2020NoGoTheoremsQuantum}, not tensor-stable~\cite{raussendorf2020PhaseSpaceSimulation}, or relying on restricted sets of free operations~\cite{thio2026KirkwoodDiracNonpositivityNecessary}. None of the previous results covered the asymptotic case with the full set of stabilizer operations. 
To prove such undistillability even with arbitrarily many copies and rounds of stabilizer operations, one would need a criterion that is stable under stabilizer operations and tensor products, and it must pick out a set of states strictly larger than the stabilizer polytope. 
This is what the PPT criterion provides in entanglement theory: LOCC cannot turn a PPT state into one with a \emph{negative partial transpose} (NPT), some mixed entangled states are PPT, and arbitrarily many tensor copies of a PPT state always remain PPT. 
Recently, progress has been made towards finding an analogue of the PPT criterion for qubit magic~\cite{liu2026TriangleCriterionMixedstate,zurel2026BasisindependentStabilizernessMaximally}, building on earlier results on magic distillability~\cite{reichardt2009QuantumUniversalityState}, but the crucial feature of tensor stability has remained elusive until now.

In this Letter, we provide the missing criterion, which is stable under tensor products and free operations for the resource theory of multi-qubit magic. 
It allows us to answer the open question of the existence of bound magic in the affirmative: Bound multi-qubit magic states exist for $n \geq 3$ qubits. This is a foundational result with practical implications for magic state distillation and quantum error correction.

The remainder of this Letter is structured as follows. 
After preliminary definitions, we motivate and define the set of multi-qubit states $\mathcal{B}_n$ and show that it is equal to the stabilizer polytope $\STAB_1$ for one qubit.
Our main theorem proves that the family of sets $\{\mathcal{B}_n\}_n$ is closed under tensor products and stabilizer operations.
We then establish the existence of bound magic by showing that $\mathcal{B}_1=\STAB_1$, while $\mathcal{B}_n$ is strictly larger than $\STAB_n$ for every $n\geq3$. For two-qubit systems, we have verified numerically that $\mathcal{B}_2=\STAB_2$.
States in $\mathcal{B}_n \setminus \STAB_n$ thus exist for $n \geq 3$ and are proven to be bound magic states by the properties of $\mathcal{B}_n$.
We provide explicit examples in the simplest three-qubit case as well as general construction methods. 
The construction of the set $\mathcal{B}_n$ uses Pauli channels which are PPT but not entanglement-breaking, which suggests a possible link between bound entanglement and bound multi-qubit magic. 
We comment on this and further follow-up questions raised by our result in the discussion.

\paragraph{Preliminaries.}
We start by introducing the notation and definitions necessary to motivate and state our result.
We work in the $n$-qubit Hilbert space $\mathcal{H}_n = (\mathbb{C}^2)^{\otimes n}$ and call $\mathcal{D}_n$ the set of density operators on $\mathcal{H}_n$.
The central mathematical objects we consider are the elements of the $n$-qubit stabilizer subtheory: Pauli operators, the Clifford group, and stabilizer states. 
We define an $n$-qubit Pauli operator $P_a$ with $a = (x,z) \in \mathbb{F}_2^n \times \mathbb{F}_2^n$ as 
\begin{equation}
    P_a = i^{x\cdot z} X^x Z^z \, ,
\end{equation}
where $X^x = \bigotimes_i X^{x_i}$ and $Z^z = \bigotimes_i Z^{z_i}$. 
The commutation relation between $n$-qubit Paulis can be written as
\begin{equation}
    P_aP_b = (-1)^{[a,b]} P_bP_a
\end{equation}
using the symplectic form $[a,b] = (x_a \cdot z_b + z_a \cdot x_b) \mod 2$.
Clifford unitaries $\Cl(n)$ on $n$ qubits form the normalizer of the Pauli group: for every $a$, conjugation by a Clifford maps $P_a$ to $\pm P_b$ for some $b$. 
Pure stabilizer states are all states that can be prepared by a Clifford unitary acting on the all-zero computational basis state.

We will also use channels constructed from Pauli operators and consider the transposition of their outputs. A Pauli channel $\mathcal{E}$ acting on a state $\rho \in \mathcal{D}_n$ is a \emph{completely positive trace-preserving} (CPTP) map of the form $\mathcal{E}(\rho) = \sum_a p_a P_a \rho P_a$, where each $P_a$ is an $n$-qubit Pauli operator and the nonnegative coefficients $p_a$, called Pauli error rates, form a probability distribution. 
The eigenvalues $\lambda_a \in [-1, 1]$ of $\mathcal{E}$ acting on Paulis, $\mathcal{E}(P_a) = \lambda_a P_a$, are the Walsh-Hadamard transform $\lambda_a = \sum_b p_b (-1)^{[a,b]}$ of the Pauli error rates and are referred to as Pauli eigenvalues.
When Pauli operators are transposed in the computational basis, each occurrence of $Y$ in a Pauli $P_a$ contributes one sign flip, so that
\begin{equation}
    P_a^T = \theta_a P_a \text{ with } \theta_a = (-1)^{\# Y \text{ in } P_a} \, .
\end{equation}

Turning to the resource theory of magic, we use its operational definition in terms of the set of free operations. These are the \textit{stabilizer operations} (SO):
\begin{definition}[Stabilizer operations]
    Following Ref.~\cite{veitch2014ResourceTheoryStabilizer}, we define stabilizer operations to be any combination of
    \begin{enumerate}
        \item computational basis state preparations,
        \item Clifford unitaries, and
        \item computational basis measurements.
    \end{enumerate}
    The choice of operation can depend adaptively on previous measurement outcomes and classical randomness. Note that this definition covers the preparation of any stabilizer state as well as any Pauli measurement.
\end{definition}

The set of free states that can be prepared using stabilizer operations is the convex hull of pure stabilizer states, which we denote by $\STAB_n$ for $n$-qubit systems. 
For a single qubit, the stabilizer polytope forms an octahedron within the Bloch sphere and can be described in terms of Pauli coordinates $r_X, r_Y, r_Z$ as $|r_X| + |r_Y| + |r_Z| \leq 1$. For multi-qubit systems, the stabilizer polytope is only fully characterized up to two qubits~\cite{reichardt2009QuantumUniversalityState} and has increasingly complicated structure as the number of qubits increases, making it an ongoing topic of research~\cite{heinrich2019RobustnessMagicSymmetries,obst2024WignersTheoremStabilizer, palhares2026TraceDistancebasedGeometric,varela2026PredictingMagicVery,liu2026GeometryTypicalityDetectability}.

\paragraph{A tensor-stable PPT analogue for multi-qubit magic.} 

The first observation that motivates our construction comes from entanglement theory: In Ref.~\cite{ruskai2003QubitEntanglementBreaking}, it has been shown that a single-qubit Pauli channel is entanglement-breaking~\cite{horodecki2003EntanglementBreakingChannels} if and only if its Pauli eigenvalues satisfy $|\lambda_X| + |\lambda_Y| + |\lambda_Z| \leq 1$. 
When such a channel acts on any single-qubit state with Bloch coordinates $r_X, r_Y, r_Z$, it is easy to see that the output will lie within the stabilizer octahedron defined above. 
Every entanglement-breaking single-qubit Pauli channel thus maps any input state $\rho$ into $\STAB_1$. 
The second observation we use is that this not only holds for entanglement-breaking Pauli channels, but also for the PPT Pauli channels, which coincide with the entanglement-breaking Pauli channels for a single qubit, but generally form a larger set for more qubits. 
Applying the notion of PPT channels~\cite{horodecki2000BindingEntanglementChannels} to Pauli channels, we define the \textit{set of PPT Pauli channels} $\PPT_n$ as follows:

\begin{definition}[PPT Pauli channels $\PPT_n$]
    Let $\mathcal{E}$ be a Pauli channel on $n$ qubits and $\mathcal{T}$ denote transposition in the computational basis. The set of PPT Pauli channels $\PPT_n$ is the set of Pauli channels for which $\mathcal{T} \circ \mathcal{E}$ is completely positive. This condition is equivalent to the Choi state of $\mathcal{E}$ being PPT. 
\end{definition}

We note that these channels have appeared in the context of entanglement theory before --- they are channels with Bell-diagonal and PPT Choi state, and have been studied in Refs.~\cite{benatti2004NondecomposableQuantumDynamical,piani2006ClassBoundEntangled}.
Expressed in coordinates $p \in \mathbb{R}^{4^n}$, the set $\PPT_n$ forms a polytope within the simplex of $4^n$-element probability distributions. 
Its cone has more recently been studied in Ref.~\cite{muller-hermes2021DecomposablePauliDiagonal}. 

The following criterion can be used to check if a given Pauli channel is PPT. It has been derived in different forms before~\cite{terhal2001HidingBitsBell,divincenzo2002QuantumDataHiding,piani2006ClassBoundEntangled,muller-hermes2021DecomposablePauliDiagonal}, but for completeness we provide a proof in the End Matter.
\begin{proposition}[PPT criterion for Pauli channels]
    \label{ppt_pauli_criterion}
Let $\mathcal{E}$ be an $n$-qubit Pauli channel with Pauli eigenvalues $\lambda_a$.
Then $\mathcal{T}\circ\mathcal{E}$ is again of Pauli form,
\begin{align}
  (\mathcal{T}\circ\mathcal{E})(\rho) &=\sum_b q_b\,P_b\,\rho\,P_b \, , \nonumber \\
  q_b &= 4^{-n}\sum_a \theta_a\lambda_a(-1)^{[a,b]} \, ,
\end{align}
with real coefficients satisfying $\sum_b q_b=1$.
The channel $\mathcal{E}$ is PPT if and only if $q_b\ge 0$ for every Pauli label $b$, that is, if and only if $\mathcal{T}\circ\mathcal{E}$ is itself a Pauli channel with Pauli error rates $q_b$.
\end{proposition}

With these definitions at hand, we can introduce the central object of this Letter: the convex hull of PPT Pauli channel outputs for arbitrary qubit numbers. 

\begin{definition}[PPT Pauli output sets $\mathcal{B}_n$]
    The set $\mathcal{B}_n$ is the convex hull of outputs of PPT Pauli channels on $n$ qubits:
    \begin{equation}
        \mathcal{B}_n = \conv\{ \mathcal{E}(\rho) : \mathcal{E} \in \PPT_n, \rho \in \mathcal{D}_n \} \, .
    \end{equation}
\end{definition}

The following theorem is the main result of this Letter and establishes that the family of sets $\{\mathcal{B}_n\}_n$ has analogous properties to the PPT states in entanglement theory and Wigner-positive states in odd-dimensional magic resource theory.

\begin{theorem}[Tensor stability and stabilizer closure of $\mathcal{B}_n$]
\label{theorem1}
    The family of sets $\{\mathcal{B}_n\}_n$ is closed under tensor products and stabilizer operations, and it reduces to the stabilizer octahedron for $n=1$ qubit.
    \begin{enumerate}
        \item[(i)] Tensor stability,
        \begin{equation}
            \rho \in \mathcal{B}_n, \sigma \in \mathcal{B}_m \Rightarrow \rho \otimes \sigma \in \mathcal{B}_{n+m} \, .
        \end{equation}
        \item[(ii)] Clifford unitaries,
                \begin{equation}
            \rho \in \mathcal{B}_n, U \in \Cl(n) \Rightarrow U\rho U^\dagger \in \mathcal{B}_n \, .
        \end{equation}
        \item[(iii)] Appending qubits,
        \begin{equation}
            \rho \in \mathcal{B}_n \Rightarrow \rho \otimes \ketbra{0}{0} \in \mathcal{B}_{n+1} \, .
        \end{equation}
        \item[(iv)] Discarding qubits ($n\geq 2$),
        \begin{equation}
              \rho \in \mathcal{B}_n \Rightarrow \Tr_n(\rho) \in \mathcal{B}_{n-1} \, .
        \end{equation}
        \item[(v)] Pauli measurements: For projectors $\Pi_\pm = \frac{1}{2} (\mathbbm{1} \pm P)$ onto the $\pm 1$ eigenspaces of any Pauli operator $P \ne \pm \mathbbm{1}$,
        \begin{equation}
            \rho \in \mathcal{B}_n \Rightarrow \Pi_\pm \rho \Pi_\pm \in \cone(\mathcal{B}_n) \, .
        \end{equation}
        \item[(vi)] Inclusion of and reduction to the free set  
        \begin{align}
            &\STAB_n \subseteq \mathcal{B}_n \text{ for all } n, \\
            \text{ and } &\mathcal{B}_1 = \STAB_1, \mathcal{B}_2 = \STAB_2 \, .
        \end{align}
    \end{enumerate}
\end{theorem}

We provide a direct proof of Theorem~\ref{theorem1} and comment on an alternative route to proving our result using Ref.~\cite{campbell2009StructureProtocolsMagic} in the End Matter.

\paragraph{Bound magic exists for $n\geq 3$ qubits.}

What remains to be shown is that $\mathcal{B}_n$ differs from $\STAB_n$ for some $n$. 
For $n=1$, the characterization of PPT Pauli channels underlying $\mathcal{B}_1 = \STAB_1$ is a reformulation of well-known results~\cite{ruskai2003QubitEntanglementBreaking,muller-hermes2021DecomposablePauliDiagonal}, although the connection to the stabilizer polytope has not previously been made explicit, to the best of our knowledge.
For $n=2$, explicit numerical verification shows that $\mathcal{B}_2 = \STAB_2$; an exact conceptual proof of this equality remains to be given.
At three qubits, however, the two sets are distinct and $\STAB_3 \subsetneq \mathcal{B}_3$. 
We will now show this by constructing a three-qubit state $\rho_B \in \mathcal{B}_3 \setminus \STAB_3$. 

The construction uses the PPT Pauli channel $\mathcal{F}$ defined as 
\begin{equation}
\label{flag_channel}
    \mathcal{F}(\rho) = \frac{1}{8} \rho + \frac{1}{24} \sum_{P \in \mathcal{P}_\mathrm{flag}} P \rho P
\end{equation}
with 
\begin{align} \nonumber
          \mathcal{P}_\mathrm{flag} = \{&IXZ, IZX, XIZ, XXZ, XZI, XZX, XZZ, \\ \nonumber
        &ZIX, ZXI, ZXX, ZXZ, ZZX, IYY, XYY, \\
        &YIY, YXY, YYI, YYX, YYZ, YZY, ZYY \} .
\end{align}

The label ``flag'' refers to a correspondence between this set of 21 Paulis and the flags of the Fano plane \cite{levay2008ThreequbitOperatorsSplit}. 
The channel $\mathcal{F}$ is in $\PPT_3$, which we can show using the criterion of Proposition~\ref{ppt_pauli_criterion}: Its eigenvalues are $1$ on $III$, $1/3$ on each of the 21 strings in $\mathcal{P}_{\mathrm{flag}}$, and $0$ on every other Pauli. Since the Paulis in $\mathcal{P}_{\mathrm{flag}}$ contain even numbers of $Y$ operators, every Pauli with nonzero $\lambda_a$ has $\theta_a = 1$, so that the coefficients $q_b$ are equal to the Pauli error rates of $\mathcal{F}$, and $\mathcal{T} \circ \mathcal{F} = \mathcal{F}$.

The input state vector needed to construct our three-qubit example is proportional to $\ket{0,0,0} + \ket{+,+,+}$. With $t = \bra{0,0,0}\ket{+,+,+} = 1/\sqrt{8}$, 
write 
\begin{equation}
    \ket{\phi_\pm} = \frac{\ket{0,0,0} \pm \ket{+,+,+}}{\sqrt{2(1 \pm t)}} \, .
\end{equation}
The three-qubit bound magic state is then given 
by
\begin{equation}
    \rho_B = \mathcal{F}(\ketbra{\phi_+}{\phi_+}) \, .
\end{equation}

An explicit magic witness as well as a short impossibility proof in the End Matter show that this state is outside $\STAB_3$. 
The membership $\rho_B \in \mathcal{B}_3$ holds by construction, as it is the 
output of a channel in $\PPT_3$. 
We have thus shown that bound magic exists in three-qubit systems, and by extension for all larger qubit numbers as well: For every $n>3$, the state $\rho_B\otimes\ketbra{0}{0}^{\otimes(n-3)}$ belongs to $\mathcal{B}_n$ by stability under appending qubits, and it is outside $\STAB_n$, since tracing out the appended qubits gives $\rho_B$. 
These are not the only examples, however. To show that not all multi-qubit bound states are reducible to the three-qubit case, we provide examples of bound magic states based on $\PPT_n$ channels for any odd $n>3$ in the End Matter.
We also note that numerically constructing further examples is easy --- one can simply search over outputs of PPT Pauli channels and test magic witnesses on them. Any such state witnessed to be magic is bound magic by our theorem.

Combining our theorem with the known result that any pure $n$-qubit nonstabilizer state can be reduced to a single-qubit nonstabilizer state \cite{reichardt2005ImprovedMagicStates,reichardt2009QuantumUniversalityState}, we can state the following corollary of our result.

\begin{corollary}[Bound magic for every $n\geq 3$]
    For any $n \geq 3$ qubits, there exist nonstabilizer states from which no pure magic can be distilled, even with arbitrary numbers of copies and rounds of stabilizer operations.
\end{corollary}

The corollary follows by contradiction: If any pure magic state could be distilled from $\mathcal{B}_n$, then Ref.~\cite{reichardt2005ImprovedMagicStates} guarantees that we could use it to distill a single-qubit magic state. 
But Theorem~\ref{theorem1} and $\mathcal{B}_1=\STAB_1$ show that distilling single-qubit magic from states in $\mathcal{B}_n$ is impossible; hence we cannot distill \textit{any} pure magic from $\mathcal{B}_n$. 
Distillation to mixed nonstabilizer states with $n\geq 3$ qubits remains a possibility, but those target states would lie in $\mathcal{B}_n$ and thus be bound magic themselves.
The impossibility also holds for asymptotic approximation: For each fixed number $k$ of output qubits, $\PPT_k$ and $\mathcal{D}_k$ are compact, so $\mathcal{B}_k$, their output set's convex hull in a finite-dimensional space, is compact. 
Every pure nonstabilizer target $\ket{\psi}$ lies outside $\mathcal{B}_k$ by the preceding argument. 
Consequently, $\max_{\tau\in\mathcal{B}_k}\bra{\psi}\tau\ket{\psi}<1$. This bound applies to every nonzero-probability branch of every stabilizer protocol, regardless of its number of input copies or rounds.

Our result could also be phrased the following way: Every PPT Pauli channel is magic-breaking~\cite{patra2024QubitMagicbreakingChannels} for $n=1$ and $n=2$; for $n\geq 3$ some are not. 
How these types of Pauli channels compare to the entanglement-breaking Pauli channels is an interesting open question. 
We were unable to prove the following conjecture, which is suggested by the equivalence at $n=1$ and numerical evidence for larger systems:
\begin{conjecture}[Entanglement-breaking Pauli outputs and the stabilizer polytope]
    \label{EB_conjecture}
    The convex hull of outputs of entanglement-breaking Pauli channels on $n$ qubits is the stabilizer polytope $\STAB_n$.
\end{conjecture}
If Conjecture~\ref{EB_conjecture} is true, it would reveal a link between entanglement and the type of multi-qubit bound magic found here --- all bound magic of this type would stem from bound entanglement in the Pauli channel's Choi state.

\paragraph{Discussion.}

We have shown that multi-qubit bound magic exists for $n\geq 3$ qubits by explicit construction of a tensor- and stabilizer-operations-stable family of states that reduces to the stabilizer octahedron for one qubit, as well as bound magic states that lie within this set but outside the stabilizer polytope. 
This establishes bound multi-qubit magic in the asymptotic sense familiar from PPT bound entanglement and Wigner-positive nonstabilizerness. 

Our result has both foundational and practical implications: Foundationally, it tells us that nonstabilizerness of multi-qubit states is not equivalent to distillability, which means that some nonstabilizer states cannot be resource states for quantum computation in the magic-state injection model. 
In practice, it has implications for quantum error correction and magic state distillation. 
We show that there exist Pauli noise channels that preserve nonstabilizerness of some inputs and under which no stabilizer code, even with syndrome postselection, yields a logical qubit outside the stabilizer octahedron; and we limit magic distillation protocols in the sense that there are nonstabilizer states which no protocol can ever turn into a pure resource state.

The kinds of noise channels and examples of bound states used in our construction are far from practically relevant regimes, but this leads directly to the first of many open questions that this result raises: While our construction is sufficient for showing the existence of multi-qubit bound magic, it does not allow us to conclude that states outside $\mathcal{B}_n$ are distillable. 
Specifically, the region of bound magic states could be larger than what we have found, and the question whether bound magic exists in single- and two-qubit systems remains open.

Beyond this question of the completeness of the bound-magic set, its existence opens up several lines of research into its properties. 
Is it possible to prove an explicit link to computational advantage by constructing efficient simulation algorithms for any stabilizer computation acting on input states from our set? 
Can the bound states be activated by a catalyst~\cite{campbell2011CatalysisActivationMagic}, and are they useful for other types of tasks, in a similar way to PPT entangled states~\cite{horodecki2005SecureKeyBound, smith2008QuantumCommunicationZeroCapacity,vertesi2014DisprovingPeresConjecture,toth2018QuantumStatesPositive}?
It would also be interesting to construct new unfaithful magic monotones that better track a state's distillability, and to consider some of the more general sets of free operations~\cite{heimendahl2022AxiomaticOperationalApproaches} that have been considered in the resource theory literature, inspired by recent results on discrimination gaps between stabilizer operations and larger sets of operations~\cite{kwon2026TestingNonstabilizernessOnly,vaglini2026CausalInequalitiesWitness}.

Finally, the connection between magic and entanglement suggested by Conjecture~\ref{EB_conjecture} and the implications of our result on various foundational results on magic are worth investigating. 
One could imagine the existence of a set of undistillable states that is simpler to describe and the membership in which could be easier to decide than the hard problem of stabilizer polytope membership~\cite{leone2026UnbearableHardnessDecidinga}, in analogy to Wigner positivity in the odd-dimensional case.
Also drawing inspiration from a result in odd dimensions showing a link between distillable magic and contextuality~\cite{howardContextualitySuppliesMagic2014,delfosse2017EquivalenceContextualityNegativity}, it could be worth revisiting the question of how magic, entanglement, and contextuality are intertwined in multi-qubit systems~\cite{bermejo-vega2017ContextualityResourceModels, raussendorf2017ContextualityWignerFunction, wright2023ContextualityCompositeSystems}.

\paragraph{Acknowledgments.} 
J.K. would like to thank Zhenhuan Liu for helpful discussions, and Lorenzo Leone, Salvatore F.E.~Oliviero and Jacopo Rizzo for 
comments on an earlier draft of this manuscript. 
J.K.~and R.K.~would also like to acknowledge the workshop on Foundations of Quantum Computing (2023, 2024, 2025) where their collaboration first developed. 
This work received support from the French government under the France 2030 investment plan, as part
of the Initiative d’Excellence d’Aix-Marseille Universit\'e-A*MIDEX, AMX-22-CEI-01. The Berlin team has been supported by the BMFTR 
(MUNIQC-Atoms, QuSol, PasQuops, Hybrid++, ADHS-IQ), the Munich Quantum Valley, the Quantum Flagship (PasQuans2, Millenion), QuantERA (HQCC,
SDPCode), the Clusters of Excellence MATH+ and ML4Q, the DFG (CRC 183, SPP 2514, BoLaCo), the Quantum Flagship (Millenion, Pasquops2),
Berlin Quantum, and the ERC (DebuQC). 

\paragraph{AI statement.} 
During the research process, Claude (Opus 5.5, Fable 5.2) and ChatGPT (5.6 Sol, 6.1 Sol) have been used to automate coding, (dis)proving of mathematical conjectures, and literature searches. 
While writing the manuscript, ChatGPT (6.1 Sol) has also been used for language editing and for checking mathematical details.

\bibliography{boundmagic}

\section{End Matter}

\subsection{Proof of Proposition~\ref{ppt_pauli_criterion}}

To prove the proposition, put $d=2^n$ and let $J_{\mathcal E}$ be the Choi state formed from the normalized maximally entangled vector $\ket{\Phi_d}$. Its Pauli expansion is
\begin{equation}
J_{\mathcal E}=d^{-2}\sum_a\lambda_a P_a\otimes P_a^T\, .
\end{equation}
Partial transposition on the output turns $J_{\mathcal{E}}$ into the Choi state of $\mathcal{T}\circ\mathcal{E}$ and multiplies each summand by $\theta_a$.
The simultaneous eigenvectors of the operators $P_a\otimes P_a^T$ are the Bell states $(P_b\otimes\mathbbm{1})\ket{\Phi_d}$, with respective eigenvalues $(-1)^{[a,b]}$. 
Thus $J_{\mathcal{T}\circ\mathcal{E}}=\sum_b q_b \ketbra{\Phi_b}{\Phi_b}$.
Since $\ketbra{\Phi_b}{\Phi_b}$ is the Choi state of $\rho\mapsto P_b\rho P_b$, the $q_b$ are the Pauli error rates of $\mathcal{T}\circ\mathcal{E}$, and this map is completely positive, \textit{i.e.}, $\mathcal{E}$ is PPT, if and only if all $q_b\ge0$.

\subsection{Proof of Theorem~\ref{theorem1}}

In this section we prove Theorem~\ref{theorem1} directly, property by property. 
We will repeatedly use the fact that any state $\rho \in \mathcal{B}_n$ is a convex combination $\rho = \sum_i c_i \mathcal{E}_i(\omega_i)$ of outputs $\mathcal{E}(\omega)$ of a PPT Pauli channel $\mathcal{E}$ acting on some input state $\omega$. 
The properties of Theorem~\ref{theorem1} involve applying linear or bilinear maps to $\rho$, so that it suffices to consider $\rho = \mathcal{E}(\omega)$ with arbitrary $\omega$ for the proofs.

\begin{itemize}
    \item[(i)] We start with the tensor product stability of $\mathcal{B}_n$. By assumption, $\rho \in \mathcal{B}_n$ and $\sigma \in \mathcal{B}_m$. By linearity, it suffices to consider $\rho = \mathcal{E}(\omega)$ and $\sigma = \mathcal{E}'(\omega')$ with PPT Pauli channels $\mathcal{E} \in \PPT_n$, $\mathcal{E}' \in \PPT_m$, and input states $\omega \in \mathcal{D}_n$, $\omega' \in \mathcal{D}_m$. This gives us
\begin{align}
    \rho \otimes \sigma &= \mathcal{E}(\omega) \otimes \mathcal{E}'(\omega')\\
    &= (\mathcal{E} \otimes \mathcal{E}') (\omega \otimes \omega') \, .
    \nonumber
\end{align}
What needs to be shown is that $\mathcal{E} \otimes \mathcal{E}'$ is in $\PPT_{n+m}$, which means that $\mathcal{E} \otimes \mathcal{E}'$ is a Pauli channel and $\mathcal{T}_{n+m} \circ (\mathcal{E} \otimes \mathcal{E}')$ is completely positive. 
First, it holds that the tensor product of two Pauli channels is still a Pauli channel. 
Secondly, transposition factorizes, so that
\begin{equation}
    \mathcal{T}_{n+m} \circ (\mathcal{E} \otimes \mathcal{E}') = (\mathcal{T}_n \circ \mathcal{E}) \otimes (\mathcal{T}_m \circ \mathcal{E}') \, .
\end{equation}
This is completely positive because it is the tensor product of two completely positive maps (by assumption). 
Thus, $\mathcal{E} \otimes \mathcal{E}'$ is in $\PPT_{n+m}$, and consequently $\rho \otimes \sigma \in \mathcal{B}_{n+m}$.

\item[(ii)] As above we write $\rho = \mathcal{E}(\omega)$ with $\mathcal{E} \in \PPT_n$ for an output in the convex combination defining $\rho$. We thus need to show that for $U \in \Cl(n)$, $U \mathcal{E}(\omega) U^\dagger \in \mathcal{B}_n$. With $\Ad_U(X) = U X U^\dagger$, we can rewrite the state as
\begin{equation}
    U \mathcal{E}(\omega) U^\dagger = (\Ad_U \circ \mathcal{E} \circ \Ad_U^{-1}) (U \omega U^\dagger) \, .
\end{equation}
We need to show that $\Ad_U \circ \mathcal{E} \circ \Ad_U^{-1} \in \PPT_n$. First, Cliffords map Paulis to Paulis and thus $\Ad_U \circ \mathcal{E} \circ \Ad_U^{-1}$ remains a Pauli channel. To show PPT, we need to commute $\mathcal{T}$ past $\Ad_U$ in $\mathcal{T} \circ \Ad_U \circ \mathcal{E} \circ \Ad_U^{-1}$. Noting that $(U X U^\dagger)^T = \bar{U} X^T \bar{U}^\dagger$ for any unitary $U$, we see that $\mathcal{T} \circ \Ad_U = \Ad_{\bar{U}} \circ \mathcal{T}$ and therefore
\begin{equation}
    \mathcal{T} \circ \Ad_U \circ \mathcal{E} \circ \Ad_U^{-1} = \Ad_{\bar{U}} \circ (\mathcal{T} \circ \mathcal{E}) \circ \Ad_U^{-1} ,
\end{equation}
which is completely positive since $\mathcal{T} \circ \mathcal{E}$ is completely positive by assumption and remains so under composition with unitary conjugations.

\item[(iii)] Upon appending a qubit in the zero-state to $\rho = \mathcal{E}(\omega)$, we get $\mathcal{E}(\omega) \otimes \ketbra{0}{0}$ and want to show its membership in $\mathcal{B}_{n+1}$. To do that, define the maximally dephasing channel on a single qubit as $\Delta(\sigma) = 1/2 (\sigma + Z\sigma Z)$, which maps $\ketbra{0}{0}$ to itself and is easily verified to be an element of $\PPT_1$. Therefore,
\begin{equation}
    \mathcal{E}(\omega) \otimes \ketbra{0}{0} = (\mathcal{E} \otimes \Delta) (\omega \otimes \ketbra{0}{0}) \, ,
\end{equation}
which is in $\mathcal{B}_{n+1}$ using the same reasoning as in (i).

\item[(iv)] Let $n\geq 2$. We split each Pauli label into its part on the first $n-1$ qubits and its part on qubit $n$, $a=(a_{<n},a_n)$, so that $P_a=P_{a_{<n}}\otimes P_{a_n}$. 
As before, it suffices to consider $\rho=\mathcal{E}(\omega)$. 
Since the partial trace over qubit $n$ is invariant under conjugation by $P_{a_n}$, discarding the last qubit gives
\begin{align}
  &\Tr_n[\mathcal{E}(\omega)]
  = \mathcal{E}_{\mathrm{red}}(\Tr_n\omega) \text{ with } \\
  &\mathcal{E}_{\mathrm{red}}(\tau)=\sum_{a_{<n}}
  \Bigl(\sum_{a_n}p_{(a_{<n},a_n)}\Bigr)P_{a_{<n}}\tau P_{a_{<n}} \, .
\end{align}
This is the Pauli channel on $n-1$ qubits whose error rates are those of $\mathcal{E}$ summed over the last qubit. 
We need to show that $\mathcal{E}_{\mathrm{red}}\in\PPT_{n-1}$. Let $\iota(\tau)=\tau\otimes \ketbra{0}{0}$ append a qubit in the zero state.
Then $\mathcal{E}_{\mathrm{red}}=\operatorname{Tr}_n\circ\,\mathcal{E}\circ\iota$, and since transposition commutes with the partial trace,
\begin{equation}
  \mathcal{T}_{n-1}\circ\mathcal{E}_{\mathrm{red}}
  = \Tr_n\circ(\mathcal{T}_n\circ\mathcal{E})\circ\iota \, .
\end{equation}
This is completely positive as a composition of completely positive maps.
Thus $\mathcal{E}_{\mathrm{red}} \in \PPT_{n-1}$ and $\Tr_n(\rho)\in\mathcal{B}_{n-1}$.

\item[(v)] For $n=1$ the assertion follows from $\STAB_1 \subseteq \mathcal{B}_1$. 
For $n\geq2$, Clifford covariance (ii) reduces the problem to measuring $Z_n$ on the $n$th qubit. 
Its projectors are $\mathbbm{1}\otimes \ketbra{s}{s}$ with $s\in\{0,1\}$, and $(\mathbbm{1}\otimes \ketbra{s}{s}) \rho (\mathbbm{1}\otimes \ketbra{s}{s}) =\langle s|_n\rho|s\rangle_n \otimes \ketbra{s}{s}$.
By (iii) and (ii), using $\ketbra{s}{s} =X^s \ketbra{0}{0} X^s$, it suffices to show $\langle s|_n\rho|s\rangle_n \in \cone(\mathcal{B}_{n-1})$, and as before we can take $\rho=\mathcal{E}(\omega)$.
A Pauli on the last qubit either keeps $\ket{s}$ up to a sign ($a_n\in\{I,Z\}$) or flips it to $\ket{s\oplus 1}$ up to a phase ($a_n\in\{X,Y\}$). 
We therefore split $\mathcal{E}_{\mathrm{red}}=\mathcal{E}_0+\mathcal{E}_1$ according to whether the last qubit is flipped,
\begin{align}
  \mathcal{E}_0(\tau)&=\sum_{a_{<n}}\bigl(p_{(a_{<n},I)}+p_{(a_{<n},Z)}\bigr) P_{a_{<n}}\tau P_{a_{<n}}\, ,\\
  \mathcal{E}_1(\tau)&=\sum_{a_{<n}}\bigl(p_{(a_{<n},X)}+p_{(a_{<n},Y)}\bigr) P_{a_{<n}}\tau P_{a_{<n}}\, .
\end{align}
Since signs and phases cancel in the expectation value, we can write
\begin{equation}
  \langle s|_n \mathcal{E}(\omega)|s\rangle_n
  = \mathcal{E}_0(\omega_s)+\mathcal{E}_1(\omega_{s\oplus1}) \, ,
  \label{zn_measurement}
\end{equation}
where each $\omega_t=\langle t|_n\omega|t\rangle_n$ is a nonnegative multiple of an $(n-1)$-qubit state.
What needs to be shown is that each nonzero $\mathcal{E}_x$ is a nonnegative multiple of a channel in $\PPT_{n-1}$. First, its error rates are nonnegative, so it is a nonnegative multiple of a Pauli channel. 
Secondly, we use the same idea as in (iv). 
Let $\iota_t(\tau)=\tau\otimes \ketbra{t}{t}$ and $\pi_s(\sigma)=\langle s|_n\sigma|s\rangle_n$. 
Applying Eq.~(\ref{zn_measurement}) to
$\omega= \tau \otimes\ketbra{s\oplus x}{s\oplus x}$ gives $\mathcal{E}_x=\pi_s\circ\mathcal{E}\circ\iota_{s\oplus x}$. 
Because $\ket{s}$ is a real vector, transposition commutes with $\pi_s$, so
\begin{equation}
  \mathcal{T}_{n-1}\circ\mathcal{E}_x
  = \pi_s\circ(\mathcal{T}_n\circ\mathcal{E})\circ\iota_{s\oplus x} \, ,
\end{equation}
which is completely positive. 
Thus both terms in Eq.~(\ref{zn_measurement}), and hence their sum, lie in $\cone(\mathcal{B}_{n-1})$, which proves $\Pi_\pm \rho \Pi_\pm \in \cone(\mathcal{B}_n)$.

\item[(vi)] The $n$-qubit all-zero state is in $\mathcal{B}_n$ because the fully dephasing channel $\Delta^{\otimes n}$ is in $\PPT_n$ and $\ketbra{0}{0}^{\otimes n} = \Delta^{\otimes n}(\ketbra{0}{0}^{\otimes n})$. By property (ii), every Clifford image of $\ketbra{0}{0}^{\otimes n}$ and thus every pure stabilizer state is in $\mathcal{B}_n$ as well, and from convexity it follows that $\STAB_n \subseteq \mathcal{B}_n$. For a single-qubit PPT Pauli channel the eigenvalues obey $|\lambda_X|+|\lambda_Y|+|\lambda_Z|\leq1$. If the input has Bloch coordinates $r_X,r_Y,r_Z$, the output coordinates are $\lambda_Xr_X,\lambda_Yr_Y,\lambda_Zr_Z$; thus $\sum_{P\in\{X,Y,Z\}}|\lambda_Pr_P|\leq1$, and every output is in $\STAB_1$. This proves $\mathcal{B}_1=\STAB_1$. The proof of $\mathcal{B}_2 = \STAB_2$ is computer-assisted: Based on the classification of the extremal rays of the cone of $\PPT_2$ given in Ref.~\cite{muller-hermes2021DecomposablePauliDiagonal}, we confirmed the equality by explicit vertex enumeration and comparison to the facets of $\STAB_2$ via the vertices of the $\Lambda$-polytope~\cite{zurel2020HiddenVariableModel}.

\end{itemize}

\subsubsection{Alternative proof using the normal form for stabilizer protocols}

We note that Ref.~\cite{campbell2009StructureProtocolsMagic} provides a shortcut for the proof. Campbell and Browne showed that any stabilizer protocol used for magic state distillation with a single-qubit target can achieve an output fidelity that is upper-bounded by the fidelity achieved with what they call a \textit{stabilizer reduction}. 
A stabilizer reduction consists of a stabilizer code projection for a code with one logical qubit, followed by decoding. 
Similarly to the proof in the previous section, one can show that for any input state $\rho^{\otimes k}$ with $\rho \in \mathcal{B}_n$ and arbitrary $k \geq 1$, all postselected, non-zero probability outcomes of a stabilizer reduction lie in $\STAB_1$. 
The result of Ref.~\cite{campbell2009StructureProtocolsMagic} then lifts this to all possible stabilizer protocols. 

\subsection{Details on the three-qubit example}

Here we show that the three-qubit state $\rho_B$ constructed in the main text is indeed nonstabilizer, $\rho_B \notin \STAB_3$. 
First, we can construct the magic witness

\begin{align}
    W = &\ketbra{0,0,0}{0,0,0} + \ketbra{+,+,+}{+,+,+} \\ \nonumber
    &- \frac{3}{2\sqrt{2}} \left(\ketbra{0,0,0}{+,+,+} + \ketbra{+,+,+}{0,0,0}\right) \, ,
\end{align}
which yields $\ev{W}{\varphi} \geq 0$ for every pure stabilizer state vector $\ket{\varphi}$, but $\Tr(\rho_B W) \approx -0.015$. 
Secondly, we show that the nonstabilizerness follows analytically from the rationality of $\STAB$.

\begin{proposition}[Nonstabilizerness of the three-qubit example]
    \label{rho_b_nonstab}
The state $\rho_B=\mathcal F(\ketbra{\phi_+}{\phi_+})$ lies outside $\STAB_3$.
\end{proposition}
\begin{proof}
Set the state vectors $\ket u$ and $\ket v$ as $\ket u=\ket{0,0,0}$, $\ket v=\ket{+,+,+}$ and $t=\braket{u}{v}=1/\sqrt8$, so that we find $\ket{\phi_\pm}=(\ket u\pm\ket v)/\sqrt{2(1\pm t)}$. 
Every Pauli $P$ in $\mathcal P_{\mathrm{flag}}$ has an even number of $Y$ factors and hence is real and symmetric; it also contains either both an $X$ and a $Z$, or two $Y$ factors. 
Therefore $\bra uP\ket u=\bra vP\ket v=0$ and $\bra uP\ket v=\bra vP\ket u$. 
It follows that $\bra{\phi_-}P\ket{\phi_+}=0$. 
The same equality holds for $P=\mathbbm1$. 
Every term in the Pauli mixture defining $\rho_B$ consequently annihilates $\ket{\phi_-}$, and $\bra{\phi_-}\rho_B\ket{\phi_-}=0$.

Now take any pure three-qubit stabilizer state $\ket{s}$. 
Its projector $\ketbra{s}{s}$ is a sum of Pauli operators with coefficients $\pm1/8$, so its matrix elements in the computational basis lie in $\mathbb{Q}(i)$. 
Since $\ket{u}$ has entries $0$ and $1$, and all entries of $\ket{v}$ equal $t$ with $t^2=1/8$,
\begin{equation}
  2(1\mp t)\,|\braket{\phi_\mp}{s}|^2 = |\braket{u}{s} \mp \braket{v}{s}|^2
  = c\mp t\,m ,
\end{equation}
with rational numbers $c=|\braket{u}{s}|^2+|\braket{v}{s}|^2$ and $m=2\operatorname{Re}(\braket{u}{s}\braket{s}{v})/t$. 
Since $t$ is irrational, $\braket{\phi_-}{s}=0$ forces $c=m=0$, and thus
$\braket{\phi_+}{s}=0$.

If $\rho_B$ were a convex mixture of pure stabilizer states, the equality $\bra{\phi_-}\rho_B\ket{\phi_-}=0$ would force every stabilizer state appearing with positive weight to be orthogonal to $\ket{\phi_-}$, and hence also to $\ket{\phi_+}$. 
This would give $\bra{\phi_+}\rho_B\ket{\phi_+}=0$. 
But the identity term of $\mathcal F$ alone contributes $1/8$ to this expectation value, while all other contributions are nonnegative. 
This is a contradiction.
\end{proof}

\vspace{3pt}
\subsection{Bound magic states for any odd qubit number $n \geq 3$}

The construction using Eq.~(\ref{flag_channel}) based on the Paulis $\mathcal{P}_\mathrm{flag}$ also generalizes to arbitrary odd qubit numbers larger than three. 
Define $\mathcal{P}_\mathrm{flag}$, for any odd $n \geq 3$, as the set of $n$-qubit Paulis which (1) contain an even number of $Y$ letters in their Pauli string, (2) are not of $Z$ type (containing only $I$ and $Z$), and (3) are not of $X$ type (containing only $I$ and $X$). 
These conditions pick out the 21 three-qubit Paulis shown in the main text, and generally a set of $(2^n - 1)(2^{n-1}-1)$ $n$-qubit Paulis. 
The $n$-qubit channel $\mathcal{F}$ that does nothing to its input with probability $2^{-n}$ and otherwise applies a uniformly random Pauli from $\mathcal{P}_\mathrm{flag}$ is still PPT, with nontrivial Pauli eigenvalues $\lambda_a = 1/(2^{n-1}-1)$ on $\mathcal{P}_\mathrm{flag}$ and zero otherwise.

The remaining parts of the argument stay the same after replacing $\ket{u} = \ket{0^n}$, $\ket{v} = \ket{+^n}$, and $t = 2^{-n/2}$ in the definition of $\ket{\phi_\pm}$. 
It still holds that for $\rho_B = \mathcal{F}(\ketbra{\phi_+}{\phi_+})$, the expectation values with respect to $\ket{\phi_\pm}$ are $\ev{\rho_B}{\phi_-} = 0$ and 
\begin{equation}
    \ev{\rho_B}{\phi_+} = \frac{2t^2}{1+t} 
    = \frac{2^{1-n}}{1+t}.
\end{equation}
Since $t^2=2^{-n}$ is rational and $t$ is irrational for every odd $n$, the proof of Proposition~\ref{rho_b_nonstab} goes through unchanged, and thus $\rho_B \notin \STAB_n$. 
We note that the requirement of odd $n$ is only a feature of these specific examples, which we chose to present because of their simplicity and the availability of an analytic proof of nonstabilizerness. 
The set $\mathcal{B}_n$ is larger than $\STAB_n$ for \textit{any} $n\geq3$, and we have numerically found examples of multi-qubit bound magic states built from $\PPT_n$ channels for even $n$ as well.

\end{document}